\documentclass[final,5p,times,twocolumn,sort&compress]{elsarticle}

\usepackage{epstopdf}
\usepackage{multirow}
\usepackage{tabularx,booktabs} 
\usepackage[hang,splitrule]{footmisc}
\usepackage{flushend}
\usepackage{fixltx2e}
\usepackage{amssymb}
\usepackage{times}

\journal{Journal of Molecular Spectroscopy}

\begin{document}

\begin{frontmatter}

\title{The B$^2\Pi-$X$^2\Pi$ electronic origin band of $^{13}$C$_6$H}

\author[1,2]{X. Bacalla}

\author[2]{D. Zhao}

\author[1,3]{E.J. Salumbides}

\author[1]{M.A. Haddad\fnref{presentadd}}
	\fntext[presentadd]{Present address: Atomic and Molecular Group, Faculty of Physics, Yazd University, Yazd, Iran.}

\author[2]{H. Linnartz}

\author[1]{W. Ubachs\corref{correspondence}}
	\cortext[correspondence]{Corresponding author.}
	\ead{w.m.g.ubachs@vu.nl}

\address[1]{Department of Physics and Astronomy, LaserLaB, VU University, De Boelelaan 1081, 1081 HV Amsterdam, The Netherlands}

\address[2]{Sackler Laboratory for Astrophysics, Leiden University, Leiden Observatory, PO Box 9513, 2300 RA Leiden, The Netherlands}

\address[3]{Department of Physics, University of San Carlos, Cebu City 6000, Philippines}

\begin{abstract}
The rotationally resolved spectrum of the B$^2\Pi-$X$^2\Pi$ electronic origin band transition of $^{13}$C$_6$H is presented. The spectrum is recorded using cavity ring-down spectroscopy in combination with supersonic plasma jets by discharging a $^{13}$C$_2$H$_2$/He/Ar gas mixture. A detailed analysis of more than a hundred fully-resolved transitions allows an accurate determination of the spectroscopic parameters for both the ground and electronically excited state of $^{13}$C$_6$H.
\end{abstract}


\end{frontmatter}

\section{Introduction}
\label{Introduction}
	The hexatriynyl radical C$_6$H, a member of the linear C$_{2n}$H series, belongs to the spectroscopically best studied carbon chain radicals. The X$^2\Pi$ ground state was studied extensively in Fourier-transform microwave (MW) studies, yielding accurate constants \cite{pea88,lin99,got10}. The laboratory data followed astronomical detections of this molecule in the cold dense cloud TMC-1 \cite{suz86,cer87}. Later this molecule was detected along other lines of sight as well, e.g. towards the carbon rich star IRC+10216 \cite{cer00}. The B$^2\Pi-$X$^2\Pi$ electronic spectrum of C$_6$H was reported in several studies. The first data were recorded in Ne matrix isolation studies after mass selective deposition \citep[see e.g.][]{fre98}. This provided approximate values for the gas phase absorptions. The first gas phase spectrum was obtained under somewhat poor experimental conditions, using a hollow cathode discharge cell \cite{kot96}. The resulting spectrum only showed unresolved features, including many bands starting from vibrationally excited states. Subsequently, high precision and rotationally resolved spectra were recorded using a new planar plasma source design \cite{mot99} that allowed the detection of rotationally cold C$_6$H and C$_6$D transients under nearly Doppler-free conditions \cite{lin99}. Combined with sensitive detection techniques, such as cavity ring-down spectroscopy, it was possible to realize excellent S/N ratios, not only for the origin band but also for transitions involving vibrationally excited levels in the upper electronic state. In combination with the available MW constants it was possible to derive precise upper state constants, for both spin-orbit components. Recently, a substantially extended study of these electronic bands, as well as new transitions involving vibronic hot bands was presented, with a focus on Renner-Teller effects in C$_6$H \cite{zha11}. In the present work we extend previous spectroscopic studies to the fully $^{13}$C-substituted isotopologue $^{13}$C$_6$H. The rotationally resolved electronic spectrum yields accurate spectroscopic constants, also including the ground state constants that have not been determined before.

\section{Experiment/method}
\label{Experiment/method}

\begin{figure*}[h!]
	\centering
	\includegraphics[trim=0cm 0.5cm 0cm 1.5cm,width=0.9\linewidth]{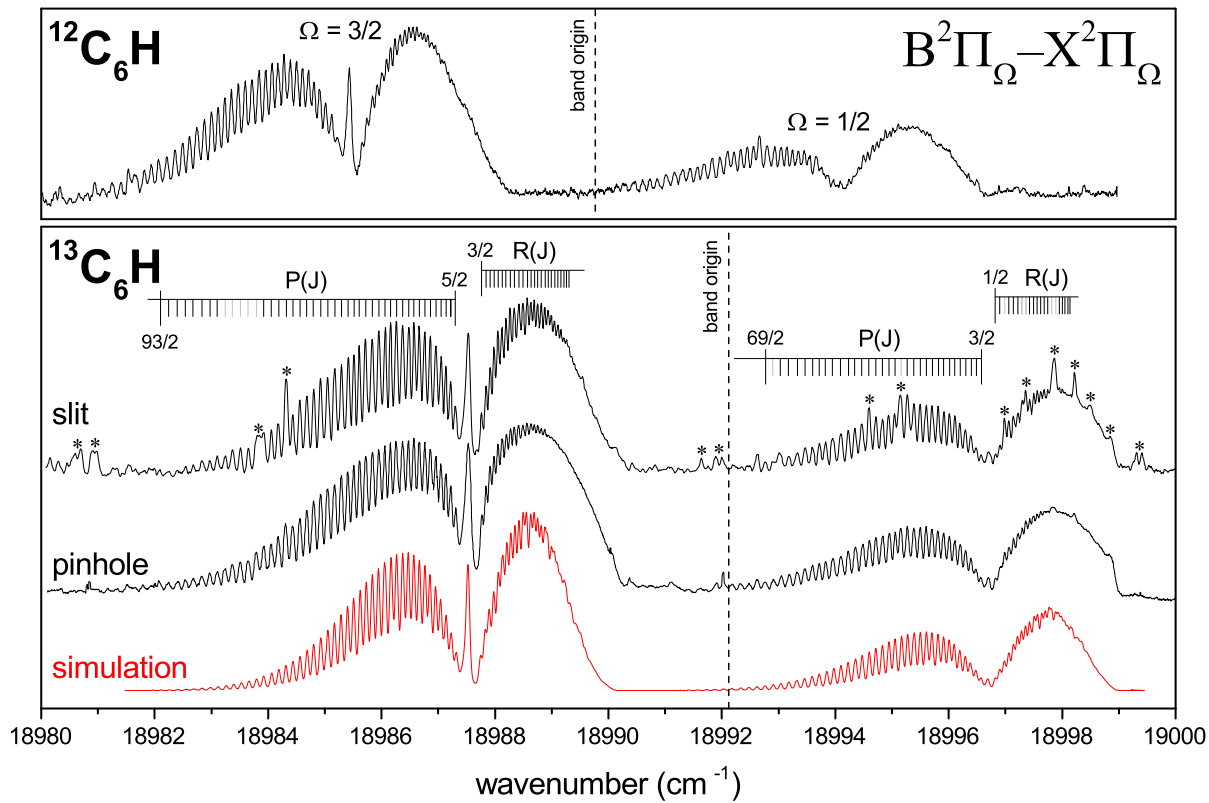}
	\caption{The B$^2\Pi-$X$^2\Pi$ electronic transition of $^{12}$C$_6$H (upper panel) and $^{13}$C$_6$H (lower panel) for a planar (slit) and a pinhole plasma expansion. Features indicated by an asterisk are due to other species and are mainly visible in the slit nozzle plasma. A simulated spectrum for $^{13}$C$_6$H is shown in the lowest trace, based on the constants listed in Table 2 and a rotational temperature of T\textsubscript{rot} = 20.3(1) K.}
\end{figure*}

	We follow an established method that has been used successfully to record electronic spectra of many different carbon chain radicals before. Details of the experimental setup and experimental procedures are available from \cite{mot99,zha11,zha11b}. Supersonically jet-cooled $^{13}$C$_6$H radicals are produced -- among other species -- in a pulsed ($\sim$10 Hz) planar or pinhole plasma expansion generated by discharging a high pressure (10 bar) gas pulse comprising of 0.18\% $^{13}$C-enriched (99\% purity) C$_2$H$_2$ in a helium:argon $\sim$85:15 mixture. Both slit and pinhole discharge nozzles, which have been used to study $^{12}$C$_6$H in Ref. \cite{zha11}, are employed in the present experiment. The 3 cm $\times$ 300 $\mu$m slit discharge nozzle, with typical discharge I/V values of $\sim$100 mA/$-$750 V per jaw, is used to generate a planar plasma expansion with a nearly Doppler-free environment and a relatively long effective absorption path length. The pinhole nozzle system, with typical  I/V $\sim$80 mA/$-$1000 V values, generates a plasma expansion suffering from a somewhat larger Doppler width, but allows measuring spectra with different rotational temperatures in a more convenient way. Final rotational temperatures typically amount to 15-30 K.

	The $^{13}$C$_6$H spectra are recorded in direct absorption using cavity ring-down laser spectroscopy. Tunable light around 525 nm -- the region where regular C$_6$H has been found -- is generated using a Nd:YAG pumped pulsed dye laser ($\sim$6 ns pulse duration). The bandwidth of the dye laser is $\sim$0.035 cm$^{-1}$, well below the expected line splitting between two adjacent rotational transitions. The light pulse is spatially filtered and enters a 58 cm long optical cavity comprising of two high reflection mirrors. The optical axis is aligned a few mm downstream, parallel to the slit or below the pinhole of the nozzle. Light leaking out of the cavity is recorded by a photo-multiplier tube and a spectrum is obtained by recording the ring-down time as function of the laser wavelength. Typical ring-down times amount to 40-100 $\mu$s. A precise pulse generator coordinates that the gas and the discharge pulse, and as well as the ring-down event, coincide. Simultaneously with the spectral recordings of $^{13}$C$_6$H, transmission fringes of an etalon (with a free spectral range of $\sim$20.1 GHz) are recorded, providing relative frequency markers to correct for a possible non-linear wavelength scanning of the dye laser. An iodine spectrum allows an absolute wavelength calibration with an accuracy better than 0.02 cm$^{-1}$.

\section{Results and analysis}
\label{Results and analysis}
	In Figure 1 the B$^2\Pi-$X$^2\Pi$ electronic origin band ($0{_0}{^0}$) transition of regular C$_6$H (upper panel) and of $^{13}$C$_6$H (lower panel) is shown. The newly recorded $^{12}$C$_6$H spectrum compares well with data presented in \cite{lin99,zha11}. From a comparison of both panels it can be concluded that the $^{13}$C-sample is highly pure, as nearly no features due to $^{12}$C contamination can be clearly seen. Some extra lines in the lower panel -- recorded through the slit expansion -- are due to overlapping hot band transitions of other small hydrocarbons in the plasma. Previous work has shown that these lines become much weaker using the pinhole discharge nozzle, as visible from the corresponding spectrum. It should be noted that in contrast to the pinhole nozzle experiments in \cite{zha11}, the final Doppler width of individual transitions is smaller, mainly because of the reduced expansion velocity of the jet that is run with more argon in the gas mixture.

	The overall spectral pattern of $^{13}$C$_6$H is very similar to that of $^{12}$C$_6$H, with a blue shift of $\sim$2 cm$^{-1}$. Both spectra consist of two components, each with (fully resolved) P, Q and (partially resolved) R branches, reflecting transitions between the two spin-orbit components: B$^2\Pi_{3/2}-$X$^2\Pi_{3/2}$ and B$^2\Pi_{1/2}-$X$^2\Pi_{1/2}$. The two band systems are split by $\Delta$A = $|$A$'-$A$''|$, the difference between the spin-orbit splitting in ground and excited state. An assignment of the individual spin-orbit components is straightforward; with a low final temperature in the jet expansion, the more intense component corresponds to a transition from the lower spin-orbit state, i.e. X$^2\Pi_{3/2}$ for C$_6$H. The observation of the barely visible Q-branch in the B$^2\Pi_{1/2}-$X$^2\Pi_{1/2}$ system provides further evidence for this assignment, as the intensity of the Q-branch scales quadratically with the value of the spin-orbit angular momenta ($\Omega^2$).

	The detailed rotational assignments for $^{13}$C$_6$H are shown in Figure 1. The determined transition energies of the 116 assigned lines listed in Table 1 are used to fit five parameters. For regular C$_6$H a full analysis has been presented before \cite{lin99,zha11}, guided by accurate ground state constants available from MW work \cite{pea88,lin99}. The focus here is on $^{13}$C$_6$H for which ground state constants are not available; it is important, therefore, to realize that ground and excited state parameters may correlate in a fit.

\begin{table*}[htbp]
  \centering
  \caption{The observed line positions (in cm$^{-1}$) for the two spin orbit components of the B$^2\Pi-$X$^2\Pi$ electronic transition of $^{13}$C$_6$H.}
    \begin{tabular}{rrrrrrrrrrrrrr}
    \toprule
    \multicolumn{2}{c}{P-branch} & \multicolumn{2}{c}{B$^2\Pi_{3/2}-$X$^2\Pi_{3/2}$} & \multicolumn{2}{c}{B$^2\Pi_{1/2}-$X$^2\Pi_{1/2}$} & \multicolumn{1}{c}{} &       & \multicolumn{2}{c}{R-branch} & \multicolumn{2}{c}{B$^2\Pi_{3/2}-$X$^2\Pi_{3/2}$} & \multicolumn{2}{c}{B$^2\Pi_{1/2}-$X$^2\Pi_{1/2}$} \\
    \cmidrule{1-6}
	\cmidrule{9-14}
    \multicolumn{1}{c}{J$'$} & \multicolumn{1}{c}{J$''$} & \multicolumn{1}{c}{Observed} & \multicolumn{1}{c}{o$-$c} & \multicolumn{1}{c}{Observed} & \multicolumn{1}{c}{o$-$c} & \multicolumn{1}{c}{} &       & \multicolumn{1}{c}{J$'$} & \multicolumn{1}{c}{J$''$} & \multicolumn{1}{c}{Observed} & \multicolumn{1}{c}{o$-$c} & \multicolumn{1}{c}{Observed} & \multicolumn{1}{c}{o$-$c} \\
    0.5   & 1.5   &       &       & 18996.57 & 0.003 &       &       & 1.5   & 0.5   &       &       & 18996.82 & $-$0.001 \\
    1.5   & 2.5   & 18987.30 & $-$0.017 & 18996.48 & 0.003 &       &       & 2.5   & 1.5   & 18987.73 & $-$0.007 & 18996.90 & 0.000 \\
    2.5   & 3.5   & 18987.22 & $-$0.010 & 18996.39 & 0.003 &       &       & 3.5   & 2.5   & 18987.82 & $-$0.007 & 18996.99 & 0.002 \\
    3.5   & 4.5   & 18987.14 & $-$0.002 & 18996.29 & 0.001 &       &       & 4.5   & 3.5   & 18987.90 & $-$0.001 & 18997.06 & $-$0.001 \\
    4.5   & 5.5   & 18987.05 & 0.008 & 18996.20 & $-$0.002 &       &       & 5.5   & 4.5   & 18987.98 & 0.003 & 18997.14 & 0.000 \\
    5.5   & 6.5   & 18986.96 & 0.006 & 18996.11 & 0.002 &       &       & 6.5   & 5.5   & 18988.06 & 0.004 & 18997.22 & 0.007 \\
    6.5   & 7.5   & 18986.86 & $-$0.002 & 18996.02 & 0.008 &       &       & 7.5   & 6.5   & 18988.13 & $-$0.001 & 18997.29 & 0.001 \\
    7.5   & 8.5   & 18986.76 & $-$0.006 & 18995.92 & 0.007 &       &       & 8.5   & 7.5   & 18988.20 & $-$0.005 & 18997.36 & $-$0.003 \\
    8.5   & 9.5   & 18986.66 & $-$0.004 & 18995.81 & 0.006 &       &       & 9.5   & 8.5   & 18988.27 & $-$0.014 & 18997.43 & $-$0.004 \\
    9.5   & 10.5  & 18986.57 & 0.002 & 18995.70 & $-$0.010 &       &       & 10.5  & 9.5   & 18988.34 & $-$0.013 & 18997.50 & 0.000 \\
    10.5  & 11.5  & 18986.48 & 0.012 & 18995.60 & $-$0.003 &       &       & 11.5  & 10.5  & 18988.42 & $-$0.003 & 18997.56 & $-$0.006 \\
    11.5  & 12.5  & 18986.38 & 0.009 & 18995.51 & 0.009 &       &       & 12.5  & 11.5  & 18988.49 & 0.002 & 18997.63 & $-$0.009 \\
    12.5  & 13.5  & 18986.26 & $-$0.004 & 18995.40 & 0.004 &       &       & 13.5  & 12.5  & 18988.57 & 0.010 & 18997.68 & $-$0.014 \\
    13.5  & 14.5  & 18986.16 & 0.000 & 18995.27 & $-$0.012 &       &       & 14.5  & 13.5  & 18988.63 & 0.005 & 18997.74 & $-$0.019 \\
    14.5  & 15.5  & 18986.06 & 0.008 &       &       &       &       & 15.5  & 14.5  & 18988.69 & 0.001 &       &  \\
    15.5  & 16.5  & 18985.95 & 0.006 & 18995.05 & $-$0.007 &       &       & 16.5  & 15.5  & 18988.75 & $-$0.001 &       &  \\
    16.5  & 17.5  & 18985.85 & 0.010 & 18994.95 & 0.006 &       &       & 17.5  & 16.5  & 18988.82 & 0.002 & 18997.95 & 0.015 \\
    17.5  & 18.5  & 18985.74 & 0.008 & 18994.83 & $-$0.001 &       &       & 18.5  & 17.5  & 18988.88 & 0.002 & 18998.00 & 0.010 \\
    18.5  & 19.5  & 18985.63 & 0.004 & 18994.71 & $-$0.005 &       &       & 19.5  & 18.5  & 18988.94 & 0.004 & 18998.04 & $-$0.004 \\
    19.5  & 20.5  & 18985.51 & 0.004 & 18994.60 & 0.000 &       &       & 20.5  & 19.5  & 18989.00 & 0.004 & 18998.10 & $-$0.002 \\
    20.5  & 21.5  & 18985.40 & 0.002 & 18994.49 & 0.010 &       &       & 21.5  & 20.5  & 18989.05 & 0.005 & 18998.15 & $-$0.004 \\
    21.5  & 22.5  & 18985.28 & $-$0.005 & 18994.37 & 0.013 &       &       & 22.5  & 21.5  & 18989.11 & 0.009 &       &  \\
    22.5  & 23.5  & 18985.16 & $-$0.005 & 18994.24 & 0.007 &       &       & 23.5  & 22.5  & 18989.16 & 0.005 &       &  \\
    23.5  & 24.5  & 18985.05 & $-$0.002 & 18994.10 & $-$0.007 &       &       & 24.5  & 23.5  & 18989.21 & 0.003 &       &  \\
    24.5  & 25.5  & 18984.93 & $-$0.002 & 18993.98 & $-$0.002 &       &       & 25.5  & 24.5  & 18989.26 & 0.001 &       &  \\
    25.5  & 26.5  & 18984.81 & 0.000 & 18993.85 & $-$0.004 &       &       & 26.5  & 25.5  & 18989.31 & $-$0.004 &       &  \\
    26.5  & 27.5  & 18984.69 & $-$0.001 & 18993.73 & 0.003 &       &       &       &       &       &       &       &  \\
    27.5  & 28.5  & 18984.56 & $-$0.008 & 18993.59 & $-$0.003 &       &       &       &       &       &       &       &  \\
    28.5  & 29.5  & 18984.44 & $-$0.004 & 18993.46 & 0.008 &       &       &       &       &       &       &       &  \\
    29.5  & 30.5  & 18984.31 & $-$0.001 & 18993.33 & 0.009 &       &       &       &       &       &       &       &  \\
    30.5  & 31.5  & 18984.18 & $-$0.006 & 18993.18 & $-$0.006 &       &       &       &       &       &       &       &  \\
    31.5  & 32.5  & 18984.05 & $-$0.004 & 18993.03 & $-$0.016 &       &       &       &       &       &       &       &  \\
    32.5  & 33.5  & 18983.92 & $-$0.005 &       &       &       &       &       &       &       &       &       &  \\
    33.5  & 34.5  &       &       & 18992.77 & 0.006 &       &       &       &       &       &       &       &  \\
    34.5  & 35.5  &       &       &       &       &       &       &       &       &       &       &       &  \\
    35.5  & 36.5  &       &       &       &       &       &       &       &       &       &       &       &  \\
    36.5  & 37.5  &       &       &       &       &       &       &       &       &       &       &       &  \\
    37.5  & 38.5  &       &       &       &       &       &       &       &       &       &       &       &  \\
    38.5  & 39.5  & 18983.10 & $-$0.013 &       &       &       &       &       &       &       &       &       &  \\
    39.5  & 40.5  & 18982.97 & 0.002 &       &       &       &       &       &       &       &       &       &  \\
    40.5  & 41.5  & 18982.83 & 0.006 &       &       &       &       &       &       &       &       &       &  \\
    41.5  & 42.5  & 18982.68 & $-$0.005 &       &       &       &       &       &       &       &       &       &  \\
    42.5  & 43.5  & 18982.54 & 0.002 &       &       &       &       &       &       &       &       &       &  \\
    43.5  & 44.5  & 18982.40 & 0.008 &       &       &       &       &       &       &       &       &       &  \\
    44.5  & 45.5  & 18982.25 & 0.001 &       &       &       &       &       &       &       &       &       &  \\
    45.5  & 46.5  & 18982.09 & 0.000 &       &       &       &       &       &       &       &       &       &  \\
    \bottomrule
    \end{tabular}
\end{table*}

	The rotational analysis is performed using \textsc{pgopher} \cite{PGOPHER} software. The transition frequencies are fitted using a standard Hamiltonian for a ${^2\Pi}-{^2\Pi}$ electronic transition, with the band origin (T$_0$), rotational parameters B$''$ and B$'$, and spin-orbit constants A$''$ and A$'$ as floating parameters. We note that the centrifugal distortion constants (D$''$ and D$'$) cannot be well determined with our data set, and therefore, their values are fixed to 1.2 $\times$ 10$^{-9}$ cm$^{-1}$. This is obtained by scaling the D$''$ value for $^{12}$C$_6$H with the reduced mass ratio ($^{12}$C$_6$H/$^{13}$C$_6$H) which is derived from the rotational constant ratio. It turns out to be possible to fit both spin-orbit components simultaneously with one set of molecular parameters, yielding an overall standard deviation of 0.007 cm$^{-1}$, i.e., well below the experimental accuracy. The resulting constants are listed under FIT-I in Table 2. The corresponding observed-calculated (o$-$c) values for individual rotational transitions are also listed in Table 1. These values exhibit a statistical behavior, indicating that the spectrum is largely free of perturbations.

\begin{table}
	\centering
	\begin{minipage}{\linewidth}
		\caption{Molecular parameters of $^{13}$C$_6$H for ground and excited ${^2\Pi}$ state derived from a line fit (FIT-I) and a combination difference fit (FIT-II), compared with values available from for $^{12}$C$_6$H \cite{lin99,zha11}. All parameters are in cm$^{-1}$.}
		\renewcommand{\arraystretch}{1.5}
		\setlength\tabcolsep{5.75pt}
		\begin{tabular}{lccc}
			\toprule
			& \multicolumn{2}{c}{$^{13}$C$_6$H} & \multirow{2}{*}{$^{12}$C$_6$H} \\
			\cmidrule{2-3}
			& FIT-I & FIT-II &  \\
			\midrule
			B$''$   & 0.042973(16) & 0.042942(17) & 0.04640497 \\
			D$''$\footnote{Parameters fixed in the least-squares fit for $^{13}$C$_6$H.\label{D}}  & $1.20 \times 10^{-9}$ & $1.20 \times 10^{-9}$ & $1.35 \times 10^{-9}$ \\
			A$''$   & $-$11.62(13) & $-$12.9(12) & $-$15.0424 \\
			B$'$    & 0.042218(17) & 0.042199(17) & 0.0455952(5) \\
			D$'$\footref{D}   & $1.20 \times 10^{-9}$ & $1.20 \times 10^{-9}$ & $1.58(28) \times 10^{-9}$ \\
			A$'$    & $-$20.78(13) & $-$26(4) & $-$23.6924(7) \\
			B$'$/B$''$\footnote{Ratio of rotational constants is dimensionless.} & 0.9824 & n/a   & 0.9825 \\
			$|$A$'-$A$''|$ & 9.16  & n/a   & 8.65 \\
			T$_0$\footnote{Values in parentheses indicate the statistical error from the least-squares fit. The absolute uncertainty in T$_0$ is limited by the uncertainty in the frequency calibration of the laser, corresponding to 0.02 cm$^{-1}$.}   & 18992.116(1) & n/a   & 18989.7672(4) \\
			\bottomrule
		\end{tabular}
		\renewcommand{\mpfootnoterule}{}
	\end{minipage}
\end{table}

	As no MW data for the ground state exists, it is wise to check for possible correlations between the resulting ground and upper state parameters using a global fit. Indeed, the fitted values for A and B give high correlation coefficients (1.000 and 0.997, respectively). For this, combination differences for both ground and excited state have been fitted separately. The resulting molecular parameters are summarized as FIT-II in Table 2. As can be seen, the derived values for (B$''$, A$''$) and (B$'$, A$'$) are in good agreement with those derived from FIT-I.

	A simulated spectrum using these constants is incorporated in the lower panel of Figure 1. The overall pattern reproduces well for a rotational temperature of T\textsubscript{rot} = 20.3(1) K. This not only applies to the Boltzmann contour, but also to the relative intensity ratio of the two spin-orbit components, indicating that the spin-orbit relaxation is as effective as the rotational cooling.

	The fitted band origin of the B$^2\Pi-$X$^2\Pi$ band of $^{13}$C$_6$H is roughly 2.35 cm$^{-1}$ blue shifted with respect to the main isotopologue. The rotational constant B$''$($^{13}$C$_6$H) is found to be 7.4\% smaller than that of B$''$($^{12}$C$_6$H), fully consistent with predictions from the previously determined C$_6$H structure in Ref. \cite{mcc05}. Upon electronic excitation, the value of B$'$ decreases, indicating that the molecule stretches, but as the inter-atomic forces are expected to barely change, the B$'$/B$''$ ratios for $^{12/13}$C$_6$H are nearly the same. It should be noted that the effective spin-orbit splitting constants ($|$A$''|$ and $|$A$'|$) for $^{13}$C$_6$H are smaller than those for $^{12}$C$_6$H. This may be due to a different Renner-Teller effect (i.e., the electronic orbital-vibration interaction) in $^{13}$C$_6$H. Indeed, spin-orbit coupling can be significantly affected by Renner-Teller effects and as discussed in Ref. \cite{zha11}, for $^{12}$C$_6$H a very strong Renner-Teller effect exists.

\section*{Acknowledgements}
This work has been supported by the Netherlands Foundation for Fundamental Research on Matter (FOM) and the Netherlands Organisation for Scientific Research (NWO) through a VICI grant and within the context of the Dutch Astrochemistry Network (DAN). X.B. and W.U. acknowledge support from the Templeton Foundation.

\end{document}